\documentclass[12pt]{article}

\usepackage{verbatim}
\usepackage{float}
\usepackage{pgfplots}
\usepackage{tikz}
\usepackage{amsmath}
\usepackage{subfigure}
\usepackage{multirow}
\usepackage{booktabs}
\usepackage[margin=1in]{geometry}
\usepackage[noadjust]{cite}

\providecommand{\keywords}[1]{\textbf{Keywords---} #1}

\pgfplotsset{compat=1.16}

\usetikzlibrary{quotes,arrows.meta}
\tikzset{
  annotated cuboid/.pic={
    \tikzset{%
      every edge quotes/.append style={midway, auto},
      /cuboid/.cd,
      #1
    }
    \draw [every edge/.append style={pic actions, densely dashed, opacity=.5}, pic actions]
    (0,0,0) coordinate (o) -- ++(-\cubescale*\cubex,0,0) coordinate (a) -- ++(0,-\cubescale*\cubey,0) coordinate (b) edge coordinate [pos=1] (g) ++(0,0,-\cubescale*\cubez)  -- ++(\cubescale*\cubex,0,0) coordinate (c) -- cycle
    (o) -- ++(0,0,-\cubescale*\cubez) coordinate (d) -- ++(0,-\cubescale*\cubey,0) coordinate (e) edge (g) -- (c) -- cycle
    (o) -- (a) -- ++(0,0,-\cubescale*\cubez) coordinate (f) edge (g) -- (d) -- cycle;
  },
  /cuboid/.search also={/tikz},
  /cuboid/.cd,
  width/.store in=\cubex,
  height/.store in=\cubey,
  depth/.store in=\cubez,
  units/.store in=\cubeunits,
  scale/.store in=\cubescale,
  width=10,
  height=10,
  depth=10,
  units=cm,
  scale=.1,
}

\date{}

\begin{document}

\title{Prediction of Hydraulic Blockage at Cross Drainage Structures using Regression Analysis}

\author{\small{Umair Iqbal\thanks{Umair Iqbal is with the SMART Infrastructure Facility, University of Wollongong, Australia, e-mail-ui010@uowmail.edu.au}, Johan Barthelemy, Pascal Perez and  Wanqing Li}
}

\maketitle

\begin{abstract}
Hydraulic blockage of cross-drainage structures such as culverts is considered one of main contributor in triggering urban flash floods. However, due to lack of during floods data and highly non-linear nature of debris interaction, conventional modelling for hydraulic blockage is not possible. This paper proposes to use machine learning regression analysis for the prediction of hydraulic blockage. Relevant data has been collected by performing a scaled in-lab study and replicating different blockage scenarios. From the regression analysis, Artificial Neural Network (ANN) was reported best in hydraulic blockage prediction with $R^2$ of 0.89. With deployment of hydraulic sensors in smart cities, and availability of Big Data, regression analysis may prove helpful in addressing the blockage detection problem which is difficult to counter using conventional experimental and hydrological approaches. 
\end{abstract}


\section{Introduction}

Blockage of cross-drainage structures by debris is highly probable in urban area and causes flash floods \cite{ARR_Report, french2015culvert, roso2004prediction, french2012non, wallerstein1996debris, blanc2013analysis}. Debris transported by the flood water accumulates across cross-drainage structures and reduces their hydraulic capacity which ultimately results in diversion of flow or overtoping \cite{Armitage2007, Ugarelli2010, Hammond2015, Santos2017}. Floods in Wollongong \cite{french2015culvert, BarthelmessMechanism, rigby2002causes, van2001modelling, davis2001analysis} and Newcastle \cite{french2015culvert, wbm2008newcastle} are highlighted events where blockage of cross-drainage structures was reported as one of the main cause for flooding. In response to these events, Australian Rainfall and Runoff (ARR) \cite{ball2016australian} initiated Project 11: Blockage of Hydraulic Structures \cite{ARR_Report} to incorporate the hydraulic blockage into design considerations. Wollongong City Council (WCC) introduced the blockage guidelines, however, were very vague and economically inefficient.  

Weeks et al. \cite{ARR_Report} defined the blockage based on the visual occlusion of opening of hydraulic structure, however, many in this domain disagreed with this definition and argued that visual blockage and hydraulic blockage are two separate concepts with no quantifiable relation among both till date \cite{french2015culvert, french2012non, french2018design, kramer2015physical}. Lack of peak floods hydraulic and visual data are considered major hinders in the study of hydraulic blockage. Janice Blanc \cite{blanc2013analysis, blanc2014analysis} and Kramer et al. \cite{kramer2015physical} proposed the idea of using scaled physical models for studying hydraulic processes. Kramer et al. \cite{kramer2015physical} was the first in this domain to propose a mathematical definition of hydraulic blockage based on the upstream water levels for blocked and unblocked conditions. Based on the proposed definition, Kramer et al. \cite{kramer2015physical} investigated the impact of different urban debris on the hydraulic blockage of culvert. However, from practical point of view, it is not easy to determine the unblocked water levels and also dependence only on upstream water level for such a complex phenomena is the over-simplification of the problem. Artificial Intelligence (AI) and machine learning have gradually became the essential part of smart cities in automating complex real-world problems \cite{barthelemy2019edge, iqbalcomputer}. Key to successful application of these approaches is availability of relevant data from network of sensors connected in smart cities i.e., Internet of Things (IoT) \cite{barthelemy2019edge, arshad2020my, arshad2019computer}. 

This paper proposes a new perspective of addressing the problem by making use of machine learning regression analysis for the prediction of hydraulic blockage. Idea of Kramer et al. \cite{kramer2015physical} was used to replicate different blockage scenarios and numerical data related to upstream water level, downstream water level, velocity, input discharge and corresponding hydraulic blockage was recorded. Three commonly used machine learning regression algorithms including $k$-Nearest Neighbour ($k$-NN), Random Forest (RF) and Artificial Neural Network (ANN) were implemented on the recorded data to investigate the potential of machine learning in predicting the hydraulic blockage. Following two are the main contributions of this paper.

\begin{enumerate}
 \item Developing a blockage dataset by performing hydraulic investigations using scaled physical models of culverts for different blockage scenarios.
\item Implementing machine learning regression algorithms to investigate the potential of machine learning in predicting the hydraulic blockage at cross-drainage structures.  
\end{enumerate}

The rest of this article is organised as follows. Section II presents information of experimental setup developed to collect the hydraulic blockage dataset for this investigation. Section III presents the background information about different machine learning regression approaches used in this investigation. Section IV presents the experimental design of the performed investigation including details about hyperparameters, validation approach and evaluation metrics. Section V presents the results and analysis about the performed regression analysis. Information about the comparative performance of different machine learning regression models for predicting hydraulic blockage is detailed under this section. Finally, Section VI presents the conclusion and important insights from the investigation.

\section{Blockage Experimental Dataset}

\begin{figure}
 \centering
 \scalebox{0.75}{\begin{tikzpicture}

\pic[fill=gray!50, text=black, draw=black] at (0,0) {annotated cuboid={width=20, height=25, depth=20}};
\pic[fill=gray!50, text=black, draw=black] at (17,0) {annotated cuboid={width=20, height=25, depth=20}};
\pic[fill=gray!20, text=black, draw=black] at (15,-0.5) {annotated cuboid={width=150, height=12, depth=20}};

\pic[fill=gray!60, text=black, draw=black] at (8.5,-0.5) {annotated cuboid={width=20, height=12, depth=20}};
\pic[fill=gray!60, text=black, draw=black, opacity=0.7] at (8.75,-0.85) {annotated cuboid={width=20, height=4, depth=10}};

\node[] at (0,1.1) {Upstream Tank};
\node[] at (16.5,1.1) {Downstream Tank};

\node[] at (5.6,0.7) {1m};
\node[] at (10.5,0.7) {1m};

\node[] at (4.3,2.2) {Upstream Point Gauge};
\node[] at (12,2.2) {Downstream Point Gauge};
\node[] at (8,2.2) {Culvert Model};

\draw[thick, ->] (4.3,2) to (4.3,-1);
\draw[thick, ->] (12,2) to (12,-1);
\draw[thick, ->] (8,2) to (8,0);

\draw[thick, <->] (12,1) to (9.3,1);
\draw[thick, <->] (7.2,1) to (4.3,1);
\end{tikzpicture}}
 \caption{Schematic of Experimental Flume Setup for Blockage Data Collection.}
 \label{fig:setup}
\end{figure}
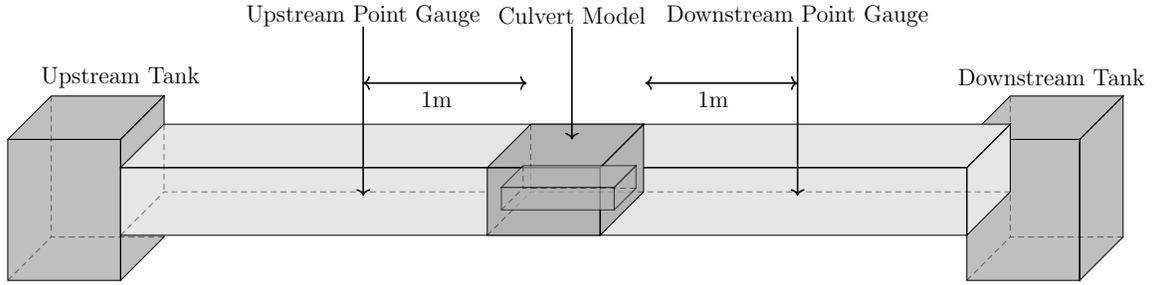

\begin{table}[]
    \centering
    \caption{Dataset Features}
    \label{tab:Dataset}
    \begin{tabular}{cl}
    \toprule
    WL\textunderscore UP & Upstream water level in blocked condition \\
    WL\textunderscore DOWN & Downstream water level in blocked condition\\
    Velocity & Upstream water velocity near culvert\\
    Culvert\textunderscore Type & Type of culvert: 0 for single circular and 1 for double circular opening\\
    Inlet\textunderscore Discharge & Inlet discharge upstream of the culvert\\
    \bottomrule
    \end{tabular}
\end{table}

\begin{figure}
    \centering
    \includegraphics[scale=0.95]{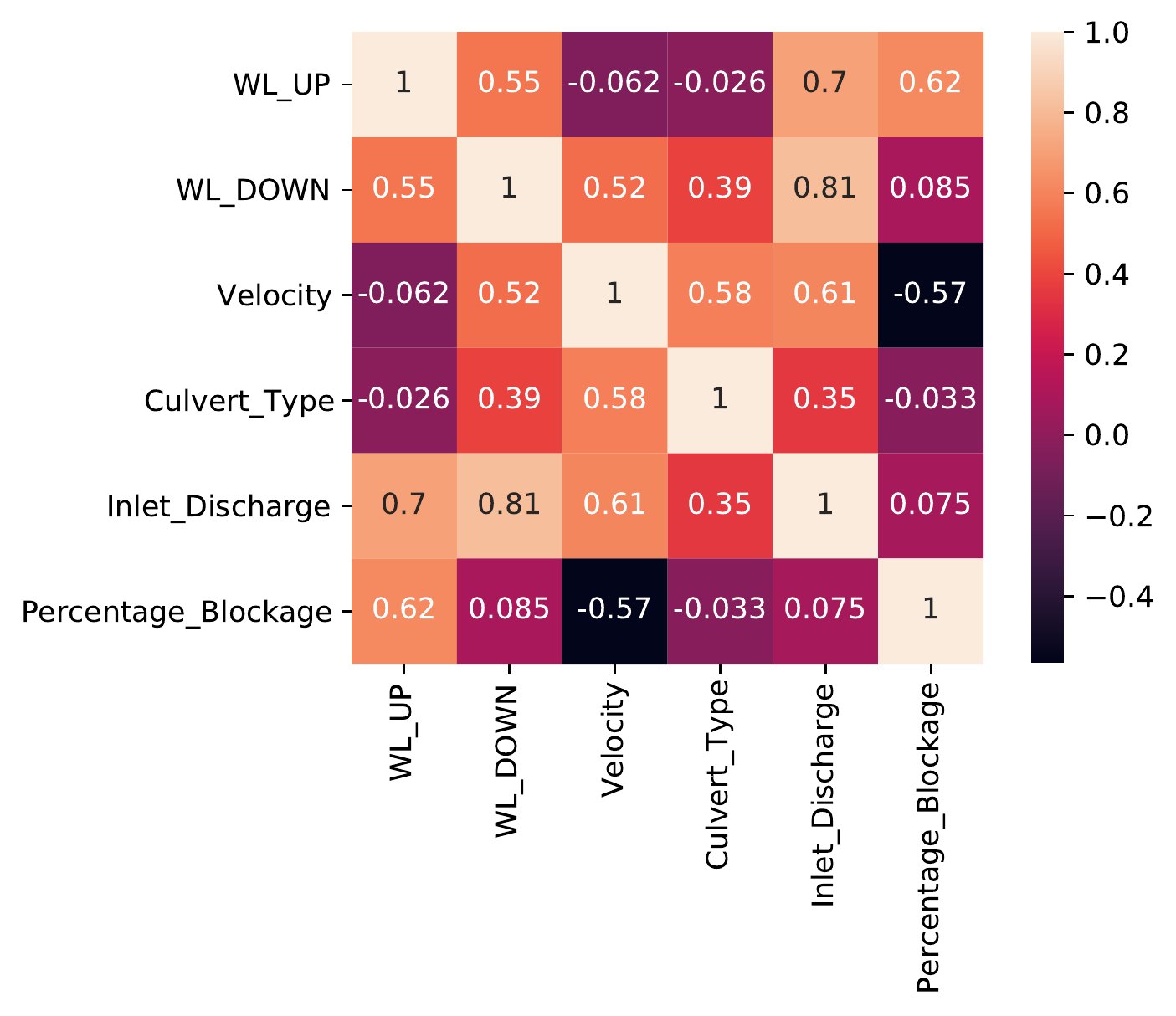}
    \caption{Correlation Heatmap for Dataset Features.}
    \label{fig:heatmap}
\end{figure}

The dataset used for this investigation was collected from a series of experiments performed at lab-scale using scaled physical models to simulate different hydraulic blockage scenarios. Experiments were performed using 12m long and 0.2m wide flume. Two different types of scaled culvert models (i.e., single circular and double circular) were used in the investigation. Flume was placed with zero slope configuration and culvert headwall height was selected as 1.1 times of opening. Water level measurements were done using a point gauge with $\pm$0.1mm accuracy. Velocity was measured using Nixon Streamflo 430 with $\pm$1.5\% accuracy. Figure \ref{fig:setup} presents the Two-Dimensional (2D) schematic of experimental setup. Hydraulic blockage scenarios were simulated using scaled urban debris (i.e., toy car, shopping car, wheelie bin, skip bin) and vegetative debris (i.e., grass, tree logs, tree branches). To simulate different flood conditions, experiments were performed for four different inlet discharge levels (i.e., 25\% submerged opening, 50\% submerged opening, 75\% submerged opening, 100\% submerged opening) taking Blanc \cite{blanc2013analysis} as reference. 

In total, 355 samples were recorded from the lab experiments for hydraulic blockage. Mathematical definition for hydraulic blockage proposed by Kramer et al. \cite{kramer2015physical} was used to calculate the blockage percentages. Although, the definition was dependent only on upstream water levels, however, it is suspected that hydraulic blockage is also dependent on other features such as inlet discharge levels, downstream water levels and upstream water velocities. Table \ref{tab:Dataset} presents the description of different features used for this study. Figure \ref{fig:heatmap} shows the correlation map between different features. It can be observed that percentage hydraulic blockage was most correlated with upstream water levels, inlet discharge and downstream water levels, respectively. 

\section{Machine Learning Approaches}

Machine learning regression algorithms aim to build a robust model based on the relationships between input features and target output. Selection of machine learning models is specific to problem, type of dataset, number of features and size of dataset. Problem of blockage prediction is a typical example of regression where a continuous variable is predicted based on five input features. For similar problem to one investigated in this article (e.g., house price prediction), $k$-NN, ANN and RF approaches have been recommended and reported to perform better \cite{engstrom2019predicting}.

\subsection{$k$-Nearest Neighbour ($k$-NN)}

$k$-NN algorithm was first introduced by Fix et al. \cite{fix1951discriminatory} in 1951 and was extended over the years by introduction of fuzzy approaches \cite{keller1985fuzzy}, weighted distance technique \cite{dudani1976distance}, refinements in Bayes error rate \cite{fukunaga1975k}, soft computing \cite{bermejo2000adaptive} and new rejection approaches \cite{hellman1970nearest}. It is a non-parametric approach in machine learning and works on the principle of predicting the target variable based on the average observations in the same neighbourhood. Euclidean distance between test and training samples is the basic idea. Given the input sample $\alpha_{\iota}=\left(\alpha_{\iota 1}, \alpha_{\iota 2}+\dots +\alpha_{\iota \rho} \right)$ with $\rho$ features and $\alpha_{\tau}$ be the test sample. Equation \ref{eq:k-nn} can be used to determine the Euclidean distance between training and test samples. 

\begin{equation} \label{eq:k-nn}
    d(\alpha_{\iota},\alpha_{\tau})=\sqrt{ \sum_{n=1}^{\rho} (\alpha_{\iota n}-\alpha_{\tau n})^{2}}
\end{equation}

Selection of appropriate value of $k$ is highly dependent on the description of problem and dataset being used. Hyperparameter optimization using algorithms (e.g., Grid Search) is a common approach to find the optimal parameters for the given dataset. Normalizing the training data improves the regression performance because of algorithm dependence on distance. 

\subsection{Random Forest (RF)}

RF is an extension of bagging technique (bootstrap aggregating) in regression (i.e., fit same regression tree repeatedly and average the results) and ensembles number of un-correlated decision trees \cite{breiman2001random}. Each tree in RF is trained over independent data sample, while samples are selected based on replacement. In simple words RF is an approach which averages multiple decision trees trained over the same training data with the aim to reduce the model variance \cite{smith2013comparison, jog2017random, li2018random}. Given a training set $(x=x_{1}+x_{2}+\dots +x_{n})$ and responses $(y=y_{1}+y_{2}+\dots+y_{n})$, bagging approach repeatedly ($m$ times) draws a training sample with replacement and fits the regression model. Once trained, predictions on the unseen data $(\hat{x})$ can be determined by taking the average of predictions of all trees on the unseen data $(\hat{x})$ \cite{breiman1996bagging}. Mathematically, it can be expressed as given in Equation \ref{eqn:RF}. 

\begin{equation}\label{eqn:RF}
    \text{Final Prediction}=\frac{1}{m}\sum_{i=1}^{m} f_{i}(\hat{x})
\end{equation}

RF algorithm extends this bagging process by selecting a random subset of features after a defined split in the learning process (i.e., feature bagging). RF is suitable for data with low bias and high variance (i.e., highly un-correlated data) \cite{breiman2001random, ao2019linear}. 

\subsection{Artificial Neural Network (ANN)}

\begin{figure}[H]
    \centering
    \begin{tikzpicture}

\node[circle, draw=black, fill=gray!30,inner sep=0pt,minimum size=20pt] (i1) at (3,6) {};
\node[circle, draw=black, fill=gray!30,inner sep=0pt,minimum size=20pt] (i2) at (3,5) {};
\node[circle, draw=black, fill=gray!30,inner sep=0pt,minimum size=20pt] (i3) at (3,4) {};
\node[]  at (3,3) {$\vdots$};
\node[circle, draw=black, fill=gray!30,inner sep=0pt,minimum size=20pt] (i4) at (3,2) {};

\node[circle, draw=black, fill=gray!30,inner sep=0pt,minimum size=20pt] (h11) at (6,8) {};
\node[circle, draw=black, fill=gray!30,inner sep=0pt,minimum size=20pt] (h12) at (6,7) {};
\node[circle, draw=black, fill=gray!30,inner sep=0pt,minimum size=20pt] (h13) at (6,6) {};
\node[circle, draw=black, fill=gray!30,inner sep=0pt,minimum size=20pt] (h14) at (6,5) {};
\node[circle, draw=black, fill=gray!30,inner sep=0pt,minimum size=20pt] (h15) at (6,4) {};
\node[circle, draw=black, fill=gray!30,inner sep=0pt,minimum size=20pt] (h16) at (6,3) {};
\node[] at (6,2) {$\vdots$};
\node[circle, draw=black, fill=gray!30,inner sep=0pt,minimum size=20pt] (h1L) at (6,1) {};

\node[circle, draw=black, fill=gray!30,inner sep=0pt,minimum size=20pt] (h21) at (9,8) {};
\node[circle, draw=black, fill=gray!30,inner sep=0pt,minimum size=20pt] (h22) at (9,7) {};
\node[circle, draw=black, fill=gray!30,inner sep=0pt,minimum size=20pt] (h23) at (9,6) {};
\node[circle, draw=black, fill=gray!30,inner sep=0pt,minimum size=20pt] (h24) at (9,5) {};
\node[circle, draw=black, fill=gray!30,inner sep=0pt,minimum size=20pt] (h25) at (9,4) {};
\node[circle, draw=black, fill=gray!30,inner sep=0pt,minimum size=20pt] (h26) at (9,3) {};
\node[] at (9,2) {$\vdots$};
\node[circle, draw=black, fill=gray!30,inner sep=0pt,minimum size=20pt] (h2L) at (9,1) {};

\node[circle, draw=black, fill=gray!30,inner sep=0pt,minimum size=20pt] (h31) at (12,8) {};
\node[circle, draw=black, fill=gray!30,inner sep=0pt,minimum size=20pt] (h32) at (12,7) {};
\node[circle, draw=black, fill=gray!30,inner sep=0pt,minimum size=20pt] (h33) at (12,6) {};
\node[circle, draw=black, fill=gray!30,inner sep=0pt,minimum size=20pt] (h34) at (12,5) {};
\node[circle, draw=black, fill=gray!30,inner sep=0pt,minimum size=20pt] (h35) at (12,4) {};
\node[circle, draw=black, fill=gray!30,inner sep=0pt,minimum size=20pt] (h36) at (12,3) {};
\node[] at (12,2) {$\vdots$};
\node[circle, draw=black, fill=gray!30,inner sep=0pt,minimum size=20pt] (h3L) at (12,1) {};

\node[circle, draw=black, fill=gray!30,inner sep=0pt,minimum size=20pt] (o) at (15,4.5) {};

\node[] (f1) at (1,6) {\Large $x_1$};
\node[] (f2) at (1,5) {\Large $x_2$};
\node[] (f3) at (1,4) {\Large $x_3$};
\node[] (f4) at (1,2) {\Large $x_5$};

\node[] (t) at (17,4.5) {\Large $y$};

\draw[thick, -, dotted] (4.5,9.5) to (4.5,-0.5);
\draw[thick, -, dotted] (13.5,9.5) to (13.5,-0.5);

\node[] at (2.5,9) {\large Input Layer};
\node[] at (9,9) {\large Hidden Layers};
\node[] at (15.5,9) {\large Output Layer};

\node[] at (6,0) {\large $n=32$};
\node[] at (9,0) {\large $n=16$};
\node[] at (12,0) {\large $n=16$};

\draw[thick, ->] (f1) to (i1);
\draw[thick, ->] (f2) to (i2);
\draw[thick, ->] (f3) to (i3);
\draw[thick, ->] (f4) to (i4);

\draw[thick, ->] (o) to (t);

\draw[thick, ->] (i1) to (h11);
\draw[thick, ->] (i1) to (h12);
\draw[thick, ->] (i1) to (h13);
\draw[thick, ->] (i1) to (h14);
\draw[thick, ->] (i1) to (h15);
\draw[thick, ->] (i1) to (h16);
\draw[thick, ->] (i1) to (h1L);

\draw[thick, ->] (i2) to (h11);
\draw[thick, ->] (i2) to (h12);
\draw[thick, ->] (i2) to (h13);
\draw[thick, ->] (i2) to (h14);
\draw[thick, ->] (i2) to (h15);
\draw[thick, ->] (i2) to (h16);
\draw[thick, ->] (i2) to (h1L);

\draw[thick, ->] (i3) to (h11);
\draw[thick, ->] (i3) to (h12);
\draw[thick, ->] (i3) to (h13);
\draw[thick, ->] (i3) to (h14);
\draw[thick, ->] (i3) to (h15);
\draw[thick, ->] (i3) to (h16);
\draw[thick, ->] (i3) to (h1L);

\draw[thick, ->] (i4) to (h11);
\draw[thick, ->] (i4) to (h12);
\draw[thick, ->] (i4) to (h13);
\draw[thick, ->] (i4) to (h14);
\draw[thick, ->] (i4) to (h15);
\draw[thick, ->] (i4) to (h16);
\draw[thick, ->] (i4) to (h1L);

\draw[thick, ->] (h11) to (h21);
\draw[thick, ->] (h11) to (h22);
\draw[thick, ->] (h11) to (h23);
\draw[thick, ->] (h11) to (h24);
\draw[thick, ->] (h11) to (h25);
\draw[thick, ->] (h11) to (h26);
\draw[thick, ->] (h11) to (h2L);

\draw[thick, ->] (h12) to (h21);
\draw[thick, ->] (h12) to (h22);
\draw[thick, ->] (h12) to (h23);
\draw[thick, ->] (h12) to (h24);
\draw[thick, ->] (h12) to (h25);
\draw[thick, ->] (h12) to (h26);
\draw[thick, ->] (h12) to (h2L);

\draw[thick, ->] (h13) to (h21);
\draw[thick, ->] (h13) to (h22);
\draw[thick, ->] (h13) to (h23);
\draw[thick, ->] (h13) to (h24);
\draw[thick, ->] (h13) to (h25);
\draw[thick, ->] (h13) to (h26);
\draw[thick, ->] (h13) to (h2L);

\draw[thick, ->] (h14) to (h21);
\draw[thick, ->] (h14) to (h22);
\draw[thick, ->] (h14) to (h23);
\draw[thick, ->] (h14) to (h24);
\draw[thick, ->] (h14) to (h25);
\draw[thick, ->] (h14) to (h26);
\draw[thick, ->] (h14) to (h2L);

\draw[thick, ->] (h15) to (h21);
\draw[thick, ->] (h15) to (h22);
\draw[thick, ->] (h15) to (h23);
\draw[thick, ->] (h15) to (h24);
\draw[thick, ->] (h15) to (h25);
\draw[thick, ->] (h15) to (h26);
\draw[thick, ->] (h15) to (h2L);

\draw[thick, ->] (h16) to (h21);
\draw[thick, ->] (h16) to (h22);
\draw[thick, ->] (h16) to (h23);
\draw[thick, ->] (h16) to (h24);
\draw[thick, ->] (h16) to (h25);
\draw[thick, ->] (h16) to (h26);
\draw[thick, ->] (h16) to (h2L);

\draw[thick, ->] (h1L) to (h21);
\draw[thick, ->] (h1L) to (h22);
\draw[thick, ->] (h1L) to (h23);
\draw[thick, ->] (h1L) to (h24);
\draw[thick, ->] (h1L) to (h25);
\draw[thick, ->] (h1L) to (h26);
\draw[thick, ->] (h1L) to (h2L);

\draw[thick, ->] (h21) to (h31);
\draw[thick, ->] (h21) to (h32);
\draw[thick, ->] (h21) to (h33);
\draw[thick, ->] (h21) to (h34);
\draw[thick, ->] (h21) to (h35);
\draw[thick, ->] (h21) to (h36);
\draw[thick, ->] (h21) to (h3L);

\draw[thick, ->] (h22) to (h31);
\draw[thick, ->] (h22) to (h32);
\draw[thick, ->] (h22) to (h33);
\draw[thick, ->] (h22) to (h34);
\draw[thick, ->] (h22) to (h35);
\draw[thick, ->] (h22) to (h36);
\draw[thick, ->] (h22) to (h3L);

\draw[thick, ->] (h23) to (h31);
\draw[thick, ->] (h23) to (h32);
\draw[thick, ->] (h23) to (h33);
\draw[thick, ->] (h23) to (h34);
\draw[thick, ->] (h23) to (h35);
\draw[thick, ->] (h23) to (h36);
\draw[thick, ->] (h23) to (h3L);

\draw[thick, ->] (h24) to (h31);
\draw[thick, ->] (h24) to (h32);
\draw[thick, ->] (h24) to (h33);
\draw[thick, ->] (h24) to (h34);
\draw[thick, ->] (h24) to (h35);
\draw[thick, ->] (h24) to (h36);
\draw[thick, ->] (h24) to (h3L);

\draw[thick, ->] (h25) to (h31);
\draw[thick, ->] (h25) to (h32);
\draw[thick, ->] (h25) to (h33);
\draw[thick, ->] (h25) to (h34);
\draw[thick, ->] (h25) to (h35);
\draw[thick, ->] (h25) to (h36);
\draw[thick, ->] (h25) to (h3L);

\draw[thick, ->] (h26) to (h31);
\draw[thick, ->] (h26) to (h32);
\draw[thick, ->] (h26) to (h33);
\draw[thick, ->] (h26) to (h34);
\draw[thick, ->] (h26) to (h35);
\draw[thick, ->] (h26) to (h36);
\draw[thick, ->] (h26) to (h3L);

\draw[thick, ->] (h2L) to (h31);
\draw[thick, ->] (h2L) to (h32);
\draw[thick, ->] (h2L) to (h33);
\draw[thick, ->] (h2L) to (h34);
\draw[thick, ->] (h2L) to (h35);
\draw[thick, ->] (h2L) to (h36);
\draw[thick, ->] (h2L) to (h3L);

\draw[thick, ->] (h31) to (o);
\draw[thick, ->] (h32) to (o);
\draw[thick, ->] (h32) to (o);
\draw[thick, ->] (h34) to (o);
\draw[thick, ->] (h35) to (o);
\draw[thick, ->] (h36) to (o);
\draw[thick, ->] (h3L) to (o);

\node[circle, draw=black, fill=gray!30,inner sep=0pt,minimum size=20pt] (n1) at (4,-2) {$x_{1}^{l}$};
\node[circle, draw=black, fill=gray!30,inner sep=0pt,minimum size=20pt] (n2) at (4,-3) {$x_{2}^{l}$};
\node[] at (4,-4) {\vdots};
\node[circle, draw=black, fill=gray!30,inner sep=0pt,minimum size=20pt] (nn) at (4,-5) {$x_{n}^{l}$};

\node[circle, draw=black, fill=none,inner sep=0pt,minimum size=80pt] at (9,-3.5) {};

\node[circle, draw=black, fill=none,inner sep=0pt,minimum size=20pt] (add) at (8.2,-3.5) {$+$};
\node[circle, draw=black, fill=none,inner sep=0pt,minimum size=20pt] (fun) at (9.7,-3.5) {$f$};

\node[] (out) at (13,-3.5) {$x_{j}^{l+1}$};

\node[] (bias) at (8.2,-5.5) {$w_{bj}^{l}$};

\node[] at (5,-2) {$w_{1j}^{l}$};
\node[] at (5,-2.8) {$w_{2j}^{l}$};
\node[] at (5,-4.2) {$w_{nj}^{l}$};

\draw[thick, ->] (n1) to (add);
\draw[thick, ->] (n2) to (add);
\draw[thick, ->] (nn) to (add);
\draw[thick, ->] (bias) to (add);
\draw[thick, ->] (add) to (fun);
\draw[thick, ->] (fun) to (out);

\draw[->, line width=0.95mm]    (h16) to[out=200,in=150] (3,-2);
\end{tikzpicture}
    \caption{ANN Structure and Functional Illustration.}
    \label{fig:ANN}
\end{figure}
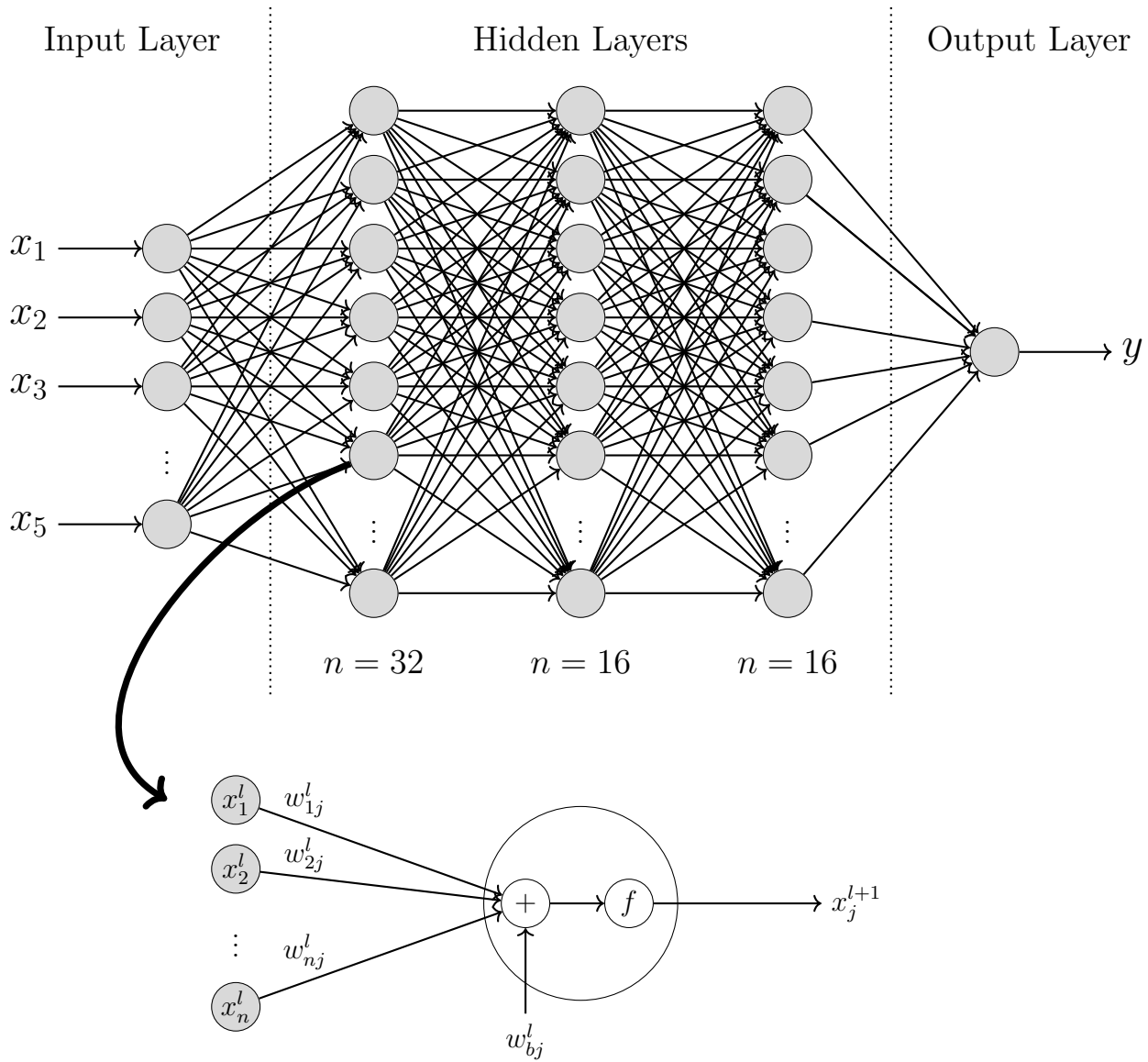

ANNs are the state of the art machine learning algorithms designed to mimic the functionality of animal brain. Nodes, connections and hidden layers are three components of network. Node in the network represents an artificial neuron which transforms the input by non-linear activation function and transmits to one or multiple neurons. Each layer in the network consists of number of artificial neurons and preforms certain transformation to input signals. Weights are assigned to each layer representing the strength of signal at neurons and are updated during the training process for improved performance. ANN consists of an input layer, an output layer and hidden layers containing artificial neurons connected with each other \cite{abraham2005artificial, mehrotra1997elements, krogh2008artificial, basheer2000artificial}. Figure \ref{fig:ANN} shows a typical representation of ANN structure used in this investigation. Where $x_1, x_{2}, \dots, x_{5}$ represent the input features and $y$ represent the target output. Figure also illustrates the functionality of single neuron in the network. It represents a neuron $j$ in layer $l+1$ which takes $x_{i}^{l}$ as input and generates $x_{j}^{l+1}$. Neuron processing can be represented mathematically as shown in Equation \ref{eqn:ANN}.

\begin{equation}\label{eqn:ANN}
x_{j}^{l+1}=f\left(\sum_{i} w_{ij}^{l}x_{i}^{l}+w_{bj}^{l}  \right)    
\end{equation}

where $w_{ij}^{l}$ denotes the weights in layer $l$, $w_{bj}^{l}$ denotes the bias term of neuron $j$ and $f$ denotes the non-linear activation function.

\section{Experimental Design}

This section presents the details about the data pre-processing, hyperparameter settings for algorithms, cross validation and evaluation metrics. 

Data was scaled using StandardScaler approach (i.e., removes the mean and scales the features to unit variance) to optimize the performance of machine learning approaches. For $k$-NN algorithm, 5 neighbours were used with leaf size of 30 and uniform weights. For RF, 100 number of trees were used with bootstrap true and MSE criterion. For ANN, three hidden layers with 32, 16 and 16 neurons were used, respectively. Furthermore, relu activation function with ``adam" solver was used. Algorithm was trained for 5000 epochs with early stopping true and learning rate of 0.001.

To establish the understanding of how well a model is trained, conventional approach is to split dataset into train and test portions and evaluate the performance on the unseen test data. However, this approach is not efficient for smaller datasets as in this case, since it reduces the number of samples used for training. Cross validation is one potential way to address this problem and is used for this investigation. Dataset was divided into 5 small folds, one used for testing and other four for training. Scores from all iterations were than averaged and presented as a better measure of model evaluation instead of conventional approach. Evaluation metrics used for assessing the performance of regression model included Root Mean Square Log Error (RMSLE), Mean Square Error (MSE), Mean Absolute Error (MAE) and $R^{2}$ score \cite{baccianella2009evaluation}. 

\begin{table}[H]
    \centering
    \caption{Summary of Empirical Results for Implemented Machine Learning Regression Algorithms}
    \label{tab:results}
    \begin{tabular}{ccccc}
    \toprule
    ~ & \textbf{MSLE} & \textbf{MSE} & \textbf{MAE} & $\mathbf{R^2}$ \\
    \toprule
$k$-Nearest Neighbour ($k$-NN)	& 0.4604	& 80.2676	& 6.8840	& 0.6672 \\
Random Forest (RF)	& 0.5524	& 106.495	& 8.0583	& 0.5584 \\
Artificial Neural Netowrk (ANN)	& \textbf{0.1931}	& \textbf{26.134}	& \textbf{3.8684}	& \textbf{0.8916} \\
\bottomrule
    \end{tabular}
\end{table}

\section{Results and Analysis} 

This section presents the results of regression models implemented for prediction of hydraulic blockage. Three different results are presented (i.e., numerical summary, scatter plots, predicted vs actual comparison) to better demonstrate the performance of models in addressing the problem.

Table \ref{tab:results} presents the summary of recorded evaluation metrics (i.e., MSLE, MSE, MAE, $R^2$) for all the implemented machine learning regression models. From the table, it can be observed that ANN outperformed all other models with $R^2$ of 0.8916. $k$-NN was the second best in terms of performance while ensemble based model (i.e., RF) was reported third best. Better performance of ANN may be attributed to the fact that model was able to better learn the features over the hidden layers. 

\begin{figure} [H]
\centering
\subfigure[K-NN]{\includegraphics[scale=0.53]{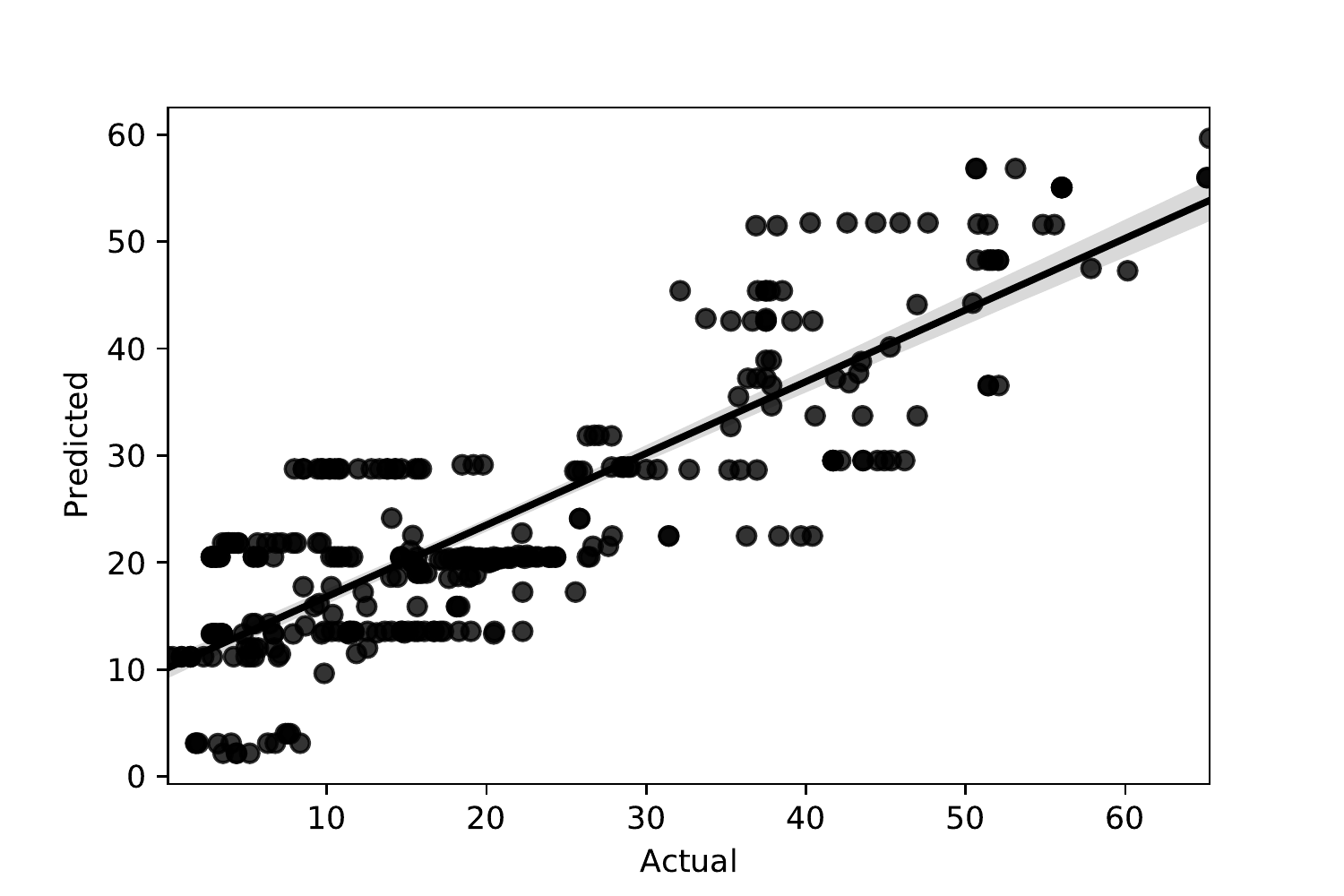}}
\subfigure[RF]{\includegraphics[scale=0.53]{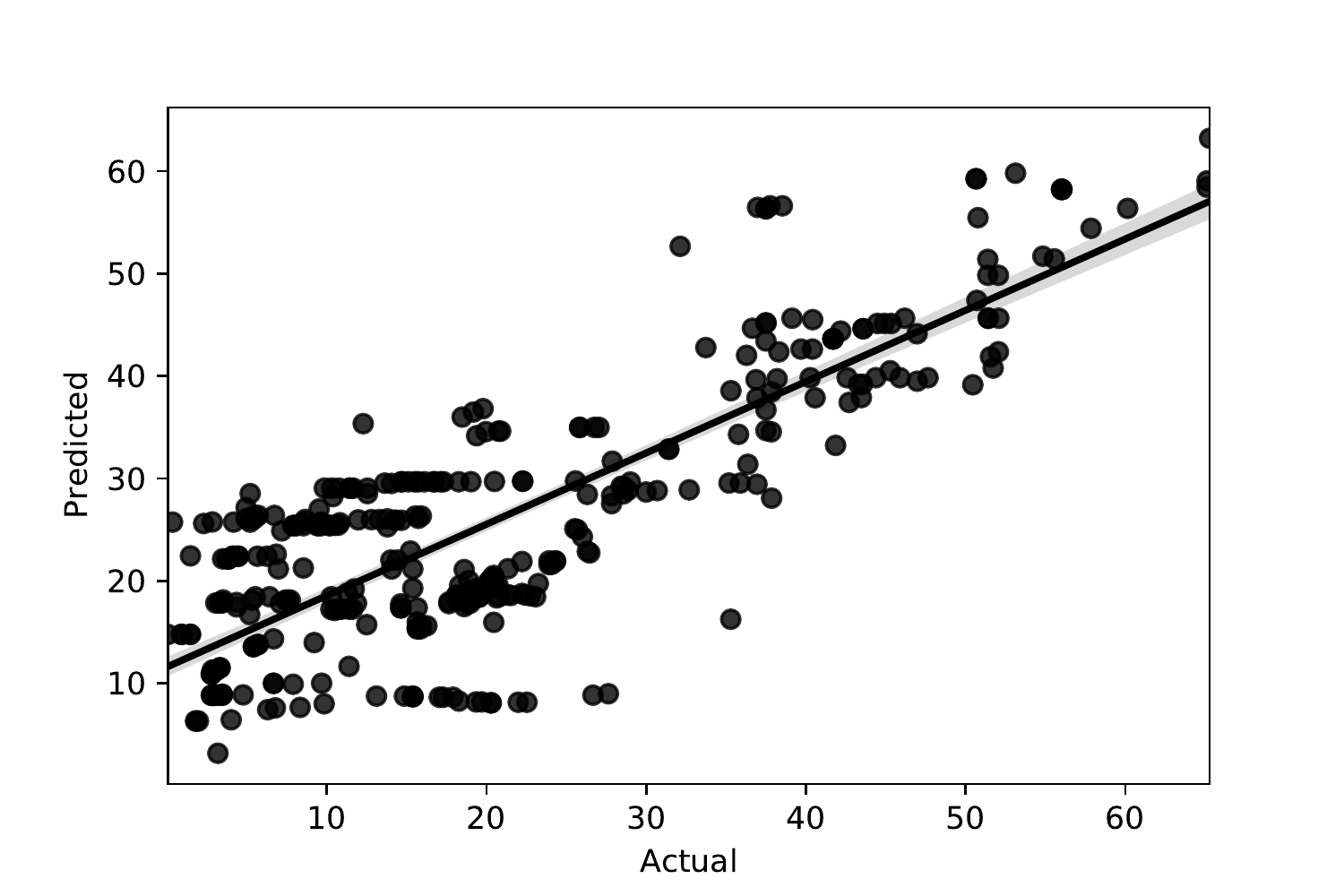}}\\
\subfigure[ANN]{\includegraphics[scale=0.53]{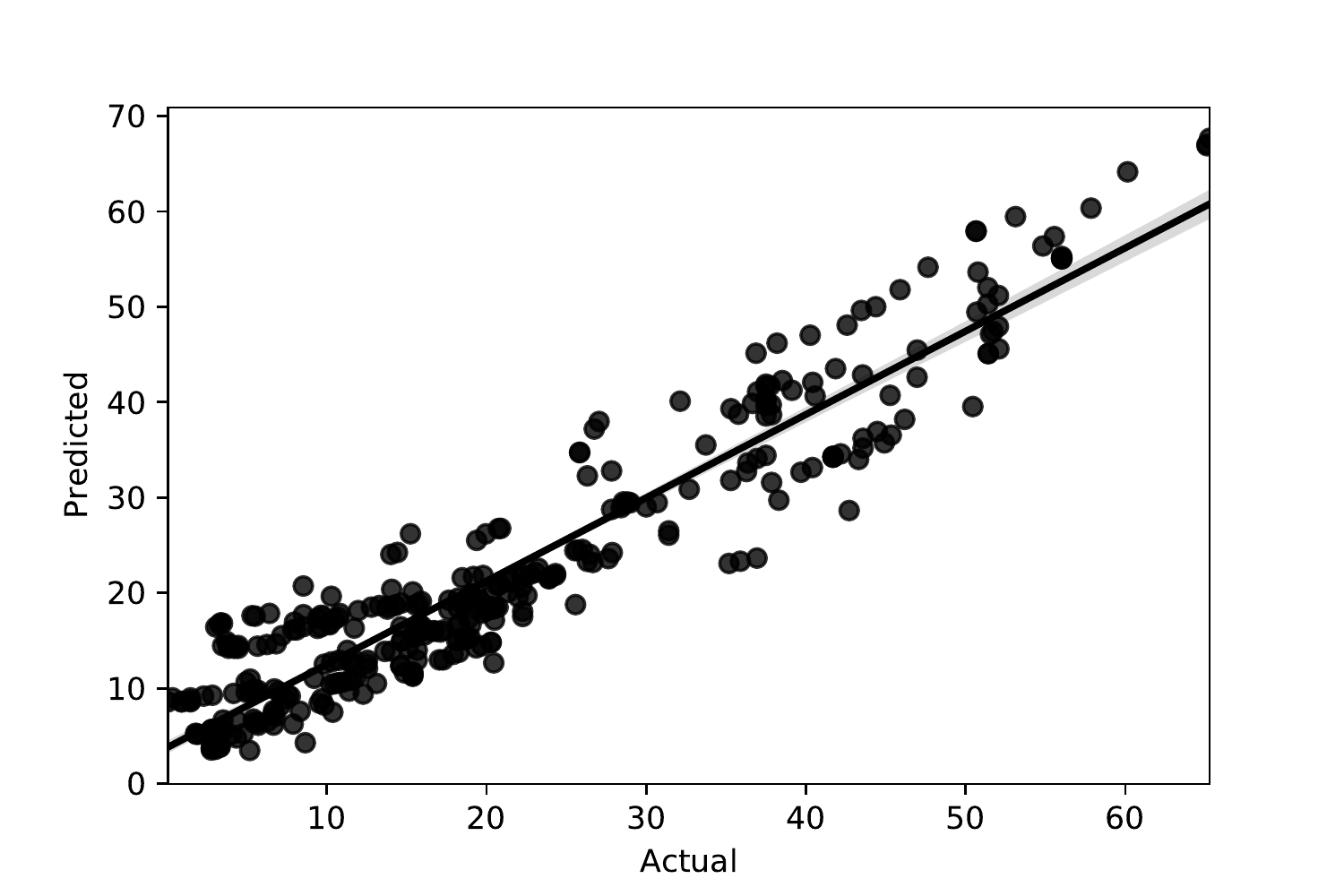}}
\caption{Scatter Plots for Implemented Regression Models to Predict Hydraulic Blockage}
\label{Scatter}
\end{figure}

\begin{figure}[H]
\centering
\subfigure[K-NN]{\includegraphics[scale=0.53]{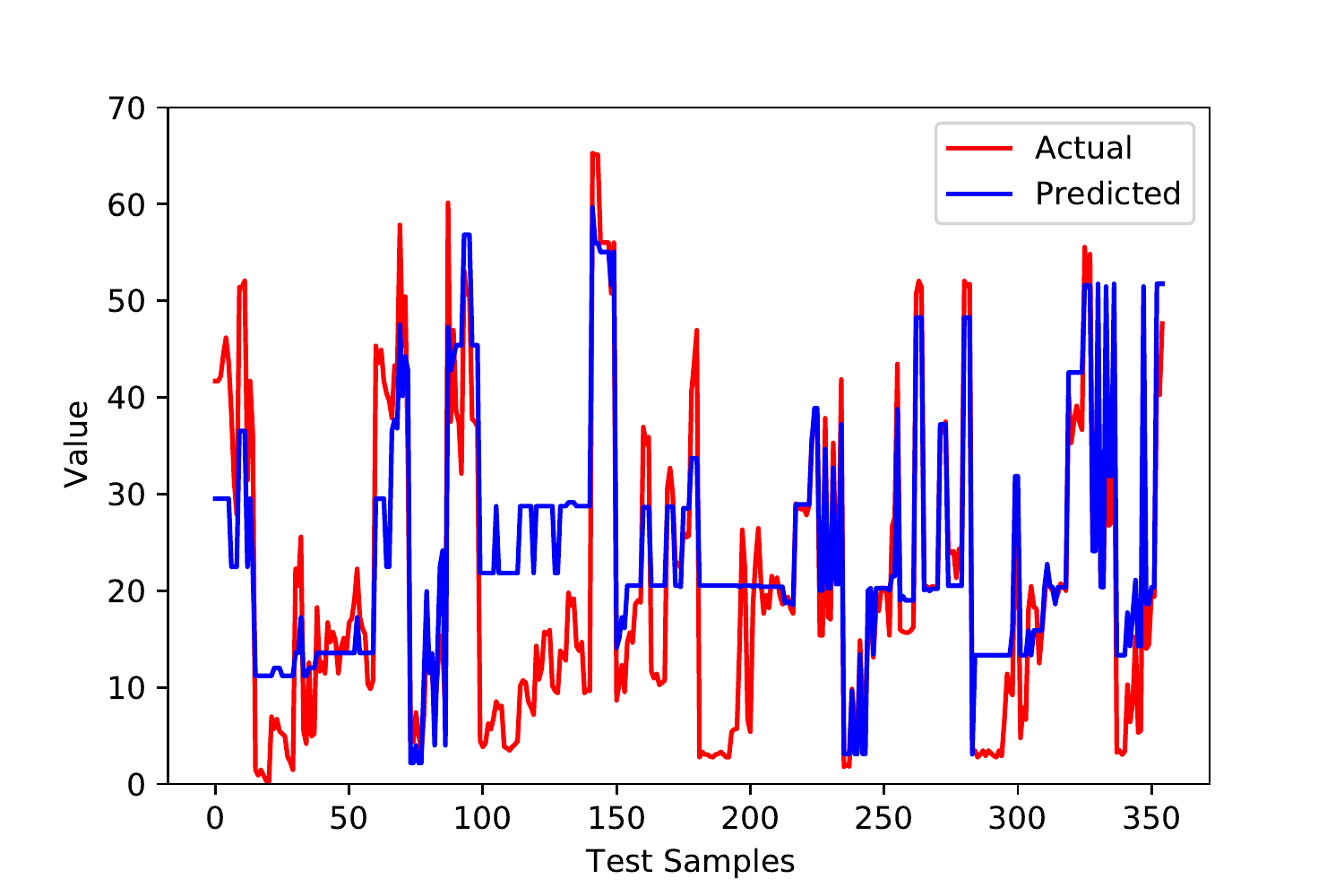}}
\subfigure[RF]{\includegraphics[scale=0.53]{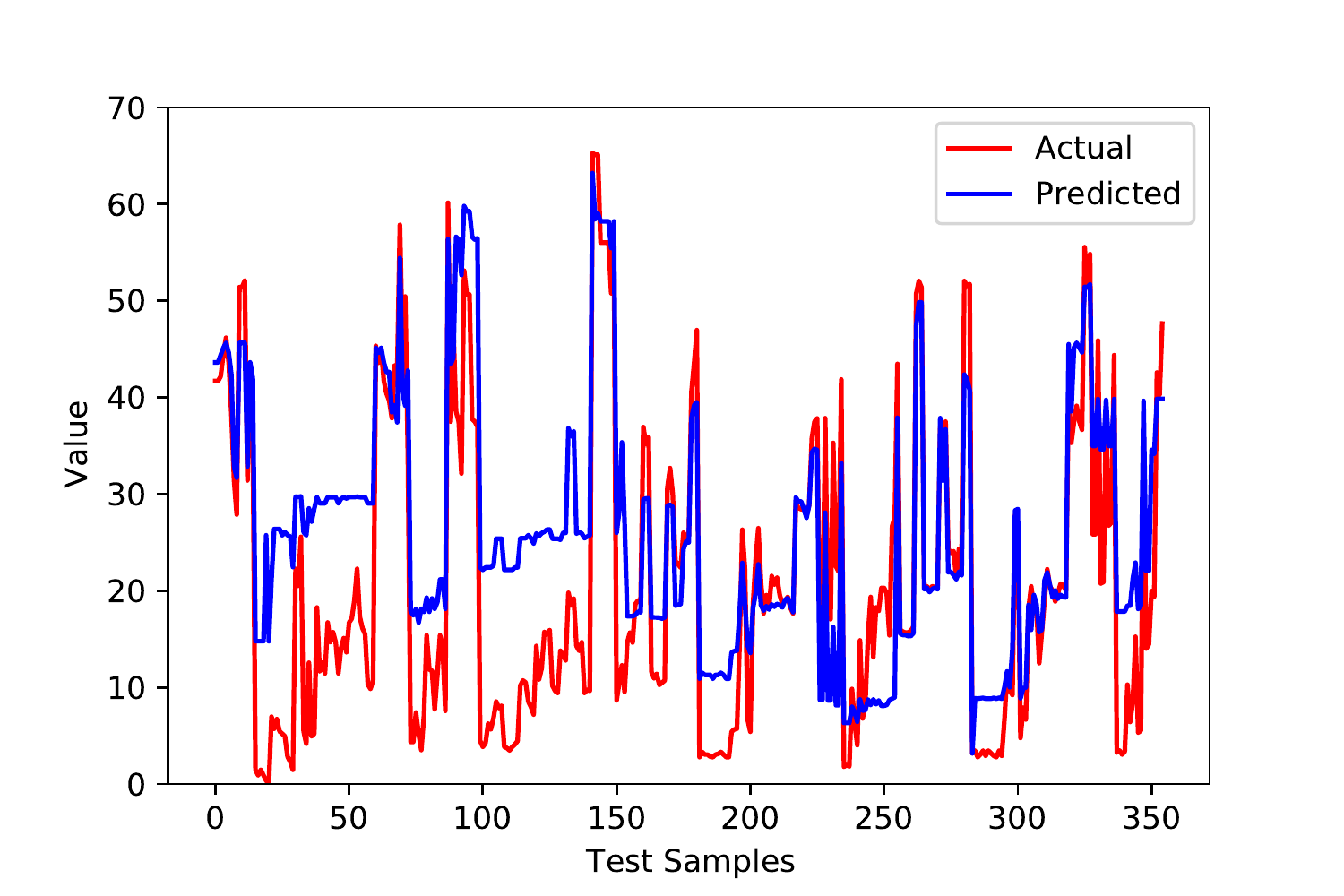}}\\
\subfigure[ANN]{\includegraphics[scale=0.53]{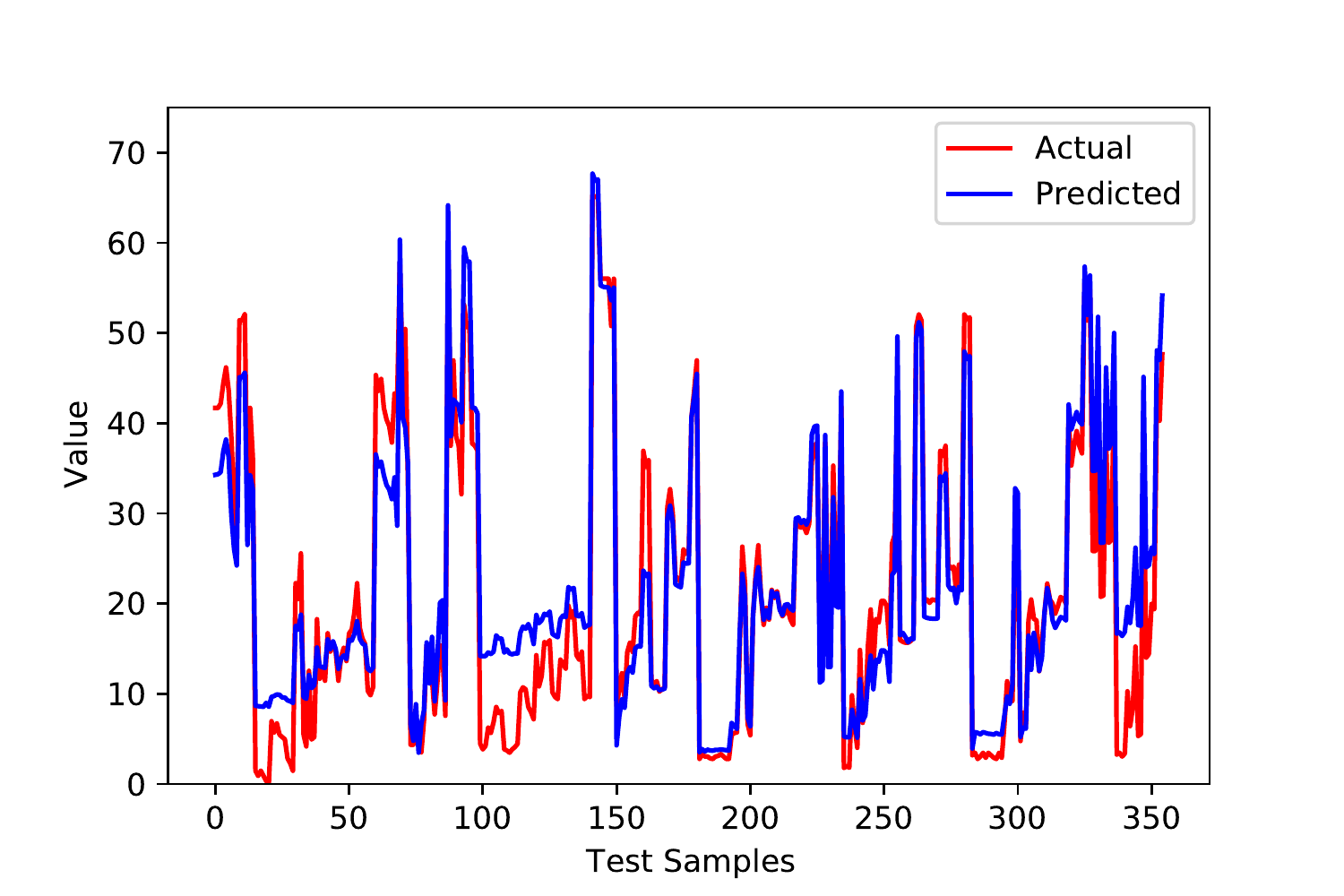}}
\caption{Actual vs Predicted Plots for Implemented Regression Models.}
\label{Prediction}
\end{figure}

Figure \ref{Scatter} shows the scatter plots for the predicted and actual values of all dataset samples to demonstrate how effectively each machine learning model was able to fit the data. It is important to mention that use of all dataset samples does not mean that models were trained and tested on the same dataset, rather, predictions are for the instances when dataset sample was in the test dataset during the cross validation process. From the Figure \ref{Scatter}, it can be clearly observed that ANN model was able to fit the data best among all other implemented models. Figure \ref{Prediction} shows the actual vs predicted plots to demonstrate how well models were able to track the actual values. From the Figure \ref{Prediction}, it can be clearly observed that for almost all cases, most samples values were under-predicted by the regression models. Degraded performance of RF can be clearly visualized by the plot. ANN model was the one with most close tracking of actual values as can be observed from the plot. 

\section{Conclusion}

Problem of hydraulic blockage at cross drainage structures was explored from a different perspective of deploying machine learning regression techniques. $k$-NN, ANN and RF techniques were successfully implemented on a dataset collected from in-lab experiments to predict the hydraulic blockage. From the analysis, ANN approach outperformed others with a decisive margin (i.e., $R^{2}=0.8916$). Growing concept of smart cities and deployment of smart sensors are the indicators for the availability of data from real events in near future. Given the complexity of modelling hydraulic blockage, analysis performed in this article suggested that regression can be a useful tool in analysing the data for hydraulic blockage prediction.

\section*{Acknowledgment}
I would like to thank the Wollongong City Council (WCC) for funding this investigation. This research was funded by the Smart Cities and Suburb Program (Round Two) of the Australian Government, grant number SCS69244. Further, I would like to thank the Higher Education Commission (HEC) of Pakistan and the University of Wollongong (UOW) for funding my PhD studies.

\bibliographystyle{IEEEtran}
\bibliography{IEEEabrv,references}

\end{document}